\documentclass[preprintnumbers, floatfix, preprintnumbers, letterpaper, superscriptaddress,nofootinbib, onecolumn]{revtex4}
\pdfoutput=1
\usepackage{graphicx,ulem}
\usepackage{amsmath}
\usepackage{amssymb}
\usepackage{subfigure}
\usepackage{hyperref}
\usepackage{url}
\usepackage{xcolor}
\usepackage{color}
\usepackage{mathrsfs}
\usepackage{calrsfs}
\usepackage{amsfonts}
\usepackage{latexsym}
\usepackage{ragged2e}
\usepackage{textcomp}
\usepackage{phaistos}
\usepackage{epstopdf}
\makeatletter
\renewcommand\@makefnmark{\hbox{\@textsuperscript{\normalfont\color{purple}\@thefnmark}}}
\renewcommand\@makefntext[1]{%
  \parindent 1em\noindent
            \hb@xt@1.8em{%
                \hss\@textsuperscript{\normalfont\@thefnmark}}#1}
\makeatother
\usepackage{caption}
\DeclareCaptionJustification{justified}{\leftskip=0pt \rightskip=0pt \parfillskip=0pt plus 1fil}
\definecolor{vividviolet}{rgb}{0.62, 0.0, 1.0}
\definecolor{amaranth}{rgb}{0.9, 0.17, 0.31}
\definecolor{palatinateblue}{rgb}{0.15, 0.23, 0.89}
\definecolor{brightpink}{rgb}{1.0, 0.0, 0.5}
\definecolor{cornflowerblue}{rgb}{0.39, 0.58, 0.93}
\definecolor{deepcarminepink}{rgb}{0.94, 0.19, 0.22}
\definecolor{radicalred}{rgb}{1.0, 0.21, 0.37}

\hypersetup{ linktoc=all,
    colorlinks, linkcolor={palatinateblue},
    citecolor={brightpink}, urlcolor={amaranth}
}

\graphicspath{{Images/}}

\graphicspath{{Images/}}

\makeatletter
\def\@fnsymbol#1{\ensuremath{\ifcase#1\or $\PHplaneTree$ \or $\textleaf$ 
\else\@ctrerr\fi}}%

\makeatother

\def\sideremark#1{\ifvmode\leavevmode\fi\vadjust{\vbox to0pt{\vss
 \hbox to 0pt{\hskip\hsize\hskip1em
 \vbox{\hsize1.5cm\tiny\raggedright\pretolerance10000
 \noindent #1\hfill}\hss}\vbox to8pt{\vfil}\vss}}}%
\def\sideremark#1{\ifvmode\leavevmode\fi\vadjust{\vbox to0pt{\vss
 \hbox to 0pt{\hskip\hsize\hskip1em
 \vbox{\hsize1.3cm\tiny\raggedright\pretolerance10000
 \noindent #1\hfill}\hss}\vbox to8pt{\vfil}\vss}}}%
\begin{document}

\title{Hairy Reissner-Nordstrom Black Holes with Asymmetric Vacua}

\author{Xiao Yan \surname{Chew}}
\email{xiao.yan.chew@just.edu.cn}
\affiliation{School of Science, Jiangsu University of Science and Technology, 212100, Zhenjiang, China}

\author{Dong-han \surname{Yeom}}
\email{innocent.yeom@gmail.com}
\affiliation{Department of Physics Education, Pusan National University, Busan 46241, Republic of Korea}
\affiliation{Research Center for Dielectric and Advanced Matter Physics, Pusan National University, Busan 46241, Republic of Korea}
\affiliation{Leung Center for Cosmology and Particle Astrophysics, National Taiwan University, Taipei 10617, Taiwan}

\begin{abstract}
We minimally coupled a scalar potential $V(\phi)$ with asymmetric vacua to the Einstein gravity to numerically construct the hairy Reissner-Nordstrom black hole (RNBH) as a direct generalization of RNBHs to possess scalar hair. By fixing the electric charge to mass ratio $q$, a branch of hairy RNBHs bifurcates from the RNBH when the scalar field $\phi_H$ is non-trivial at the horizon. The values of $q$ are bounded for $0 \leq q \leq 1$, which contrast to a class of hairy black holes with $q>1$ in the Einstein-Maxwell-scalar theory. We find that the profiles of solutions affected by the competition between the strength of $\phi_H$ and $q$, for instance, the gradient of scalar field at the horizon can increase very sharply when $q \rightarrow 1$ and $\phi_H$ is small but its gradient can be very small which independent of $q$ when $\phi_H$ is large. Furthermore, the weak energy condition of hairy RNBHs, particularly at the horizon can be satisfied when $q>0$.    
\end{abstract}

\maketitle

\section{Introduction}

According to the no-hair theorem \cite{Bekenstein:1995un,Israel:1967wq,Ruffini:1971bza}, the state of black holes in general relativity (GR) can only be described by the three global charges which are the mass, electrical charge, and angular momentum. The Reissner-Nordstrom black hole (RNBH) \cite{Reissner,Nordstrom} is the solution to the Einstein-Maxwell (EM) theory and satisfies the no-hair theorem. Nevertheless, a black hole is known as a hairy black hole when it is supported by a matter field outside the event horizon and may possess additional global charge (refers to ``hair") which is associated with the matter field. Hairy black holes can exhibit a deviation of their properties from the electrovacuum black holes in the strong gravity regime but are indistinguishable in the weak gravity regime. A mechanism which is known as the spontaneous scalarization (SS) to allow black holes can evade the no-hair theorem to possess a non-trivial scalar field $\phi$ outside the horizon, hence the RN black hole can be extended to a broader class of charged hairy black holes, for instance, a various form of scalar function $f(\phi)$ non-minimally couples with the Maxwell field \cite{Gibbons:1987ps,Garfinkle:1990qj,Dobiasch:1981vh,Gibbons:1985ac,Kallosh:1992ii,Anabalon:2013qua,Herdeiro:2018wub,Fernandes:2019rez,Myung:2018vug,Myung:2018jvi,Myung:2019oua,Astefanesei:2019mds,Astefanesei:2019pfq,Blazquez-Salcedo:2019nwd,Astefanesei:2019qsg,Myung:2020dqt,Fernandes:2020gay,Astefanesei:2020xvn,LuisBlazquez-Salcedo:2020rqp,Blazquez-Salcedo:2020nhs,Lai:2022spn,Liu:2022fxy,Kiorpelidi:2023jjw,Belkhadria:2023ooc,Guo:2023mda,Xiong:2023bpl,Meng:2024puu} in the Einstein-Maxwell-scalar (EMs) theory can give rise to the tachyonic instabilities so that charged hairy black holes can be spontaneously scalarized from the RNBH. Few decades ago T. Damour and G. Esposito-Far\`ese \cite{Damour:1993hw} have proposed the concept of SS to predict the deviation of properties for the neutron stars from GR in the strong gravity regime but becomes indistinguishable in the weak gravity regime within the framework of the scalar-tensor theory, which non-minimally couples a scalar function with the Ricci scalar.

However, one may overlook that an RNBH can also be extended to another class of charged hairy black holes in the simplest and direct manner, i.e., one can minimally couple the Einstein gravity and Maxwell field with a scalar potential $V(\phi)$. By properly introducing the form of $V(\phi)$ which is associated with the profile of $V(\phi)$ to evade the no-hair theorem, the solutions of hairy black holes can be bifurcated from the electrovacuum black holes and they are regular everywhere in the spacetime. Recently, the authors have employed two different profiles of $V(\phi)$ to construct the neutral hairy black holes without the anticipation of other extended objects, for instance the Gauss-Bonnet term or matter fields and study their properties in detail \cite{Corichi:2005pa,Chew:2022enh,Chew:2023olq}. The first profile of $V(\phi)$ with asymmetric vacua which contain a local maximum, a local minimum, and a global minimum to describe the phase transition of vacuum bubbles from the false vacuum (local minimum) to the true vacuum (global minimum) \cite{Coleman:1980aw}. The second profile of $V(\phi)$ with the shape of the inverted Mexican hat contains two degenerate maxima and a local minimum \cite{Chew:2023olq}. Therefore, we generalize those neutral hairy black holes in \cite{Chew:2022enh} to possess an electric charge and study their properties in this paper. Other choices of $V(\phi)$ or similar construction can refer to \cite{Bechmann:1995sa,Dennhardt:1996cz,Bronnikov:2001ah,Martinez:2004nb,Nikonov:2008zz,Anabalon:2012ih,Lan:2023cvz,Ahn:2014fwa,Lee:2018zym}.  

On the other hand, the RNBH possesses an inner horizon, instead of the outer horizon. The inner horizon is known as the Cauchy horizon which can preserve the predictability in GR. Recently, various attempts have been made to prove the non-existence of Cauchy horizon for the static and charged black holes \cite{An:2021plu,Cai:2020wrp,Devecioglu:2021xug,Chew:2023upu,Ong:2023lbr}. Therefore, it provides a good motivation for us to construct such charged black holes as a first step to studying the existence of the Cauchy horizon.  

This paper is organized as follows. In Sec.~\ref{sec:th}, we briefly introduce our theoretical setup comprising the Lagrangian and the metric ansatz. Then, we derive the set of coupled differential equations and study the asymptotic behavior of the functions. In Sec.~\ref{sec:prop}, we briefly introduce the quantities of interest for the black holes. In Sec~\ref{sec:res}, we present and discuss our numerical findings. Finally, in Sec.~\ref{sec:con}, we summarize our work and present an outlook.

\section{Theoretical Setting} \label{sec:th}

\subsection{Theory and Ans\"atze}

\begin{figure}
\centering
 \includegraphics[trim=50mm 170mm 20mm 20mm,scale=0.58]{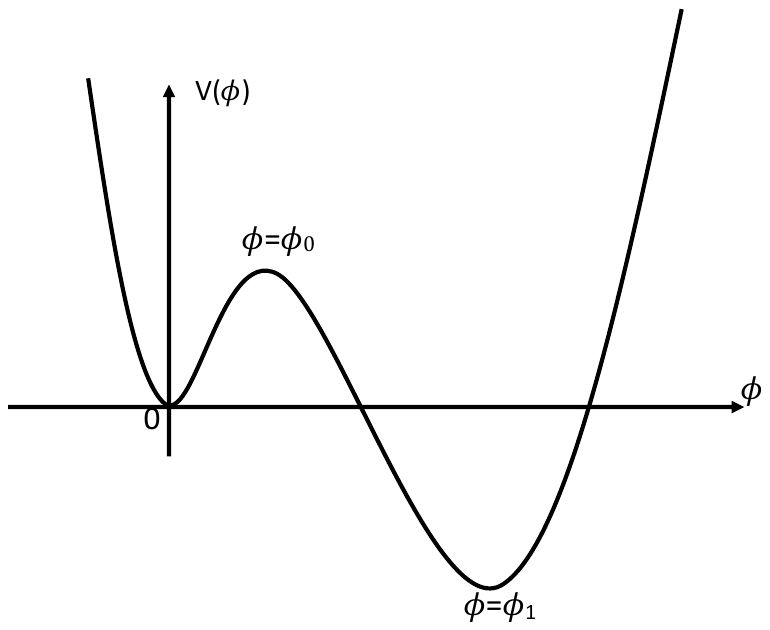}
\caption{The authors have considered the scalar potential $V(\phi)$ with a false vacuum  at $\phi=0$, a barrier at $\phi=\phi_0$ and a true vacuum at $\phi=\phi_1$ to construct the hairy black holes \cite{Chew:2022enh}.}
\label{plot_Vphi}
\end{figure}

In the Einstein-Maxwell-Klein-Gordon (EMKG) system, we consider an asymmetric potential $V(\phi)$ of a scalar field $\phi$ which is given by \cite{Corichi:2005pa,Chew:2022enh} to minimally couple with the Maxwell field and the Einstein gravity,
\begin{equation} \label{EHaction}
 S=  \int d^4 x \sqrt{-g}  \left[  \frac{R}{16 \pi G} - \frac{1}{4} F_{\mu \nu} F^{\mu \nu}-  \frac{1}{2} \partial_\mu \phi \partial^\mu \phi - V(\phi) \right]  \,,
\end{equation}
where $F_{\mu \nu}=\partial_\mu A_\nu-\partial_\nu A_\mu$ is the electromagnetic field strength tensor. The explicit form of $V(\phi)$ is given by
\begin{equation} \label{Vpot}
 V(\phi) = \frac{V_0}{12} \left( \phi - a \right)^2 \left[ 3 \left( \phi-a\right)^2 - 4 (\phi-a) (\phi_0 + \phi_1) + 6 \phi_0 \phi_1  \right] \,,
\end{equation}
with $a$, $V_0$, $\phi_0$, and $\phi_1$ are the constants. As shown in Fig.~\ref{plot_Vphi}, when $\phi=a$, this potential $V(\phi)$ possesses a local minimum which is also a zero of $V(\phi)$. The asymptotic value of $\phi$ at the infinity is fixed by $\phi=a$. $V(\phi)$ also possesses a local maximum at $\phi=a+\phi_0$ and a global minimum at $\phi=a+\phi_1$. In this paper, we choose $a=0$ such that the scalar field is asymptotically flat at the spatial infinity. Note that the asymmetrical profile of $V(\phi)$ is caused by the appearance of cubic term $\phi^3$, thus $V(\phi)$ can become symmetric if the cubic term disappears, for instance, the authors recently have employed a symmetric profile of $V(\phi)$ which possesses two degenerate global maxima and local minimum to construct the hairy black holes \cite{Chew:2023olq}. Furthermore, a similar form of Eq.~\eqref{Vpot} has been applied to construct the Fermionic star \cite{DelGrosso:2023dmv}.

Then we obtain the Einstein equation, Klein-Gordon equation, and Maxwell equation by varying the action Eq.~\eqref{EHaction} with respect to the metric, scalar field, and Maxwell field, respectively
\begin{align} 
 R_{\mu \nu} - \frac{1}{2} g_{\mu \nu} R &=  \beta  T_{\mu \nu} \,,  \label{einstein_eqn} \\
 \nabla_\mu \nabla^\mu \phi  &=  \frac{d V}{d \phi} \,,  \label{KGeqn}  \\
 \nabla_\mu F^{\mu \nu} &=0 \,, \label{Maxwell_eqn}
\end{align}
where  $\beta=8 \pi G$ and the stress-energy tensor $T_{\mu \nu}$ is given by
\begin{equation}
 T_{\mu \nu} =  -g_{\mu \nu} \left(  \frac{1}{2} \partial_\alpha \phi \partial^\alpha \phi + V(\phi) \right) + \partial_\mu \phi \partial_\nu \phi +   g^{\alpha \beta} F_{\mu \alpha} F_{\nu \beta} - \frac{1}{4} g_{\mu \nu} F_{\alpha \beta} F^{\alpha \beta} \,.
\end{equation}

We employ the following spherically symmetric ansatz to construct the charged hairy black hole solutions,
\begin{equation}  \label{line_element}
ds^2 = - N(r) e^{-2 \sigma(r)} dt^2 + \frac{dr^2}{N(r)} + r^2  \left( d \theta^2+\sin^2 \theta d\varphi^2 \right) \,, 
\end{equation}
where $N(r)=1-2 m(r)/r$ with $m(r)$ is the Misner-Sharp mass function \cite{Misner:1964je}. Note that we can read off the mass of black holes with the condition $m(\infty)=M$ where $M$ is the Arnowitt-Deser-Misner (ADM) mass. Moreover, the ansatz for the gauge field is chosen to be
\begin{equation}
 A_\mu = U(r) dt \,.
\end{equation}

\subsection{Ordinary Differential Equations (ODEs)}

To derive the ordinary differential equations (ODEs) from the Eqs.~\eqref{einstein_eqn}-\eqref{Maxwell_eqn}, we begin with Eq. \eqref{Maxwell_eqn} to directly obtain a first-order ODE, 
\begin{equation}
 U' = \frac{Q e^{-\sigma}}{r^2} \,, \label{Maxwell_ode}
\end{equation}
where the prime denotes the derivative of the functions with respect to the radial coordinate $r$ and we denote $Q$ as the electric charge. The substitution of Eq. \eqref{line_element} into the EMKG system yields a set of nonlinear ODEs for the following functions,
\begin{equation}
m' = \frac{\beta}{4} r^2 \left( N \phi'^2 + 2 V  + \frac{Q^2}{r^4} \right) \,, \quad \sigma' = - \frac{\beta}{2} r \phi'^2 \,,    \quad
\left(  e^{- \sigma} r^2 N \phi' \right)' = e^{- \sigma} r^2  \frac{d V}{d \phi} \,.
\end{equation}
If the scalar field vanishes $(\phi=0)$, then the trivial solution for the EMKG system is the RNBH which is given by
\begin{equation}
    m(r) = M - \frac{Q^2}{2 r} \,, \quad \sigma(r) = 0 \,, \quad U(r) = U_{\infty} - \frac{Q}{r} \,,
\end{equation}
where $M$ is the ADM mass and $U_{\infty}$ is the electric potential. The horizon $r_H$ of the RNBH is given by $r_H=M+\sqrt{M^2-Q^2}$. However, if the scalar field doesn't vanish but the scalar potential vanishes $(V(\phi)=0)$, the solution of the above EMKG system is still the RNBH but with the scalar field diverges at the horizon. Hence, the proper introduction of $V(\phi)$ is very crucial so that it can regularize the scalar field at the horizon. Previously the authors have considered Eq.~\eqref{Vpot} to construct the neutral hairy black holes and study their properties. Thus, in this paper, we generalize them to possess the electric charge $Q$.

To construct the charged hairy black hole solutions that are globally regular, the functions and their derivatives are required to be finite, particularly at the horizon. Hence, the asymptotic behavior of the functions at the horizon can be described by the power series expansions which the few leading terms in the series expansion are given by
\begin{align}
 m(r) &= \frac{r_H}{2}+ m_1 (r-r_H) + O\left( (r-r_H)^2 \right) \,, \\
\sigma(r) &= \sigma_H + \sigma_1   (r-r_H) + O\left( (r-r_H)^2 \right)  \,, \label{m_ex} \\
 \phi(r) &= \phi_H +  \phi_{H,1}  (r-r_H) + O\left( (r-r_H)^2 \right)  \,, \\
U(r) &= \frac{Q e^{-\sigma_H}}{r^2_H}  (r-r_H)  -  \frac{ Q e^{-\sigma_H} \left( 2 + \sigma_1 r_H   \right) }{2 r^3_H}  (r-r_H)^2   + O\left( (r-r_H)^3 \right)  \,,
\end{align} 
where
\begin{equation}
   m_1 = \frac{\beta}{4} r^2_H \left(  2  V(\phi_H) + \frac{ Q^2}{ r^4_H}  \right)  \,, \quad  \sigma_1 = -  \frac{\beta r_H}{2} \phi^2_{H,1} \,, \quad   \phi_{H,1}= \frac{r_H \frac{d V(\phi_H)}{d \phi}}{1-\beta r_H^2 V(\phi_H)-\frac{\beta Q^2}{2 r^2_H}}  \,. \label{hor}
\end{equation} 
Here $\sigma_H$ and $\phi_H$ are the values of $\sigma$ and $\phi$ at the horizon. The asymptotic expansion for the functions at infinity, which are given by
\begin{align}
    m(r) &= M  + \tilde{m}_1 \frac{\exp{(- 2 m_{\text{eff}} r)}}{r} + ...\, , \\
    \sigma(r) &= \tilde{\sigma}_1 \frac{ \exp{(- 2 m_{\text{eff}} r)}}{r}  +   ... \, , \\
    \phi(r) &= \tilde{\phi}_{H,1}  \frac{ \exp{(- m_{\text{eff}} r)} }{r} + ... \, , \\
     U(r) &= U_\infty -  \frac{Q}{r}  + ... \,,
\end{align}
where $\tilde{m}_1$, $\tilde{\sigma}_1$ and $\tilde{\phi}_{H,1}$ are constants; $U_\infty$ is the electric potential and $M$ is the ADM mass of the charged hairy black holes. Note that the denominator of $\phi_{H,1}$ has to be imposed with the condition $1-\beta r_H^2 V(\phi_H)-\frac{\beta Q^2}{2 r^2_H} \neq 0$ in order to keep $\phi(r)$ and $\sigma(r)$ finite at the horizon. Moreover, the effective mass of the scalar field is given by $m_{\text{eff}}=\sqrt{V_0 \phi_0 \phi_1}$.

Since the ODEs are nonlinear, it would be very challenging to obtain the closed form for the charged hairy black holes, although the ODEs look very simple, so we integrate the ODEs by the professional solver Colsys which employs the Newton-Raphson method to solve a set of nonlinear ODEs by providing the adaptive mesh refinement to generate the solutions with high accuracy and the estimation of errors of solutions \cite{Ascher:1979iha}. In the numerics we compactify the radial coordinate $r$ by $r=r_H/(1-x)$ with $x \in [0,1]$ which can map the one-to-one correspondence of horizon and infinity to $0$ and $1$, respectively. We also introduce the following dimensionless parameters, 
\begin{equation}
 r \rightarrow  \frac{r}{\sqrt{\beta} } \,, \quad m \rightarrow  \frac{m}{\sqrt{\beta} } \,, \quad \phi \rightarrow \sqrt{\beta} \phi \,, \quad \phi_1 \rightarrow \sqrt{\beta} \phi_1 \,, \quad \phi_0 \rightarrow \sqrt{\beta} \phi_0 \,, \quad V \rightarrow \sqrt{\beta} V  \,.
\end{equation}
Therefore, we are left with the following free parameters: $\phi_0$, $\phi_1$, $r_H$, $q$, $\sigma_H$, $\phi_H$,  $\tilde{m}_1$, $\tilde{\sigma}_1$, $\tilde{\phi}_{H,1}$, $U_\infty$ and $M$. The parameters $\sigma_H$, $\tilde{m}_1$, $\tilde{\sigma}_1$, $\tilde{\phi}_{H,1}$, $U_\infty$ and $M$ are determined when the solutions satisfy the boundary conditions, thus the input parameters in the numerics are given by $\phi_0$, $\phi_1$, $r_H$, $q$ and $\phi_H$.

\subsection{Basic Properties of Charged Hairy Black Holes} \label{sec:prop}

In this subsection, we study the basic properties of charged hairy black holes, in particular, we are interested in the area of horizon $A_H$ and Hawking temperature $T_H$ of black holes,
\begin{equation}
A_H = 4 \pi r^2_H \,, \quad   T_H = \frac{1}{4 \pi} N'(r_H) e^{-\sigma_H} \,, 
\end{equation}
where $\sigma_H$ is the value of function $\sigma$ at the horizon. For the convenience of comparing our black hole solution with a known solution, which is the RNBH in this case, we introduce the following reduced quantities at the horizon of the black holes,
\begin{equation}
 a_H = \frac{A_H}{16 \pi M^2} \,, \quad t_H = 8 \pi T_H M \,.
\end{equation}
The explicit form of $a_H$ and $t_H$ for the RNBH are given by \cite{Astefanesei:2019pfq}
\begin{equation}
 a_H = \frac{1}{4} \left( 1 + \sqrt{1-q^2}  \right)^2 \,, \quad t_H = \frac{4 \sqrt{1-q^2}}{\left( 1+\sqrt{1-q^2} \right)^2} \,,    
\end{equation}
where $q$ is interpreted as ratio of the charge $Q$ to the ADM mass $M$ and defined as
\begin{equation}
    q = \frac{Q}{M} \,.
\end{equation}
The values of $a_H$ and $t_H$ are unity when the black hole is Schwarzchild black hole $(q=0)$. $a_H$ is bounded in between $1/4 \leq a_H \leq 1$ for RN black hole where $a_H=1/4$ corresponds to the extremal RNBH $(q=1)$. Moreover, $t_H$ is bounded in between $0 \leq t_H \leq 1$ for the RNBH where $t_H=0$ corresponds to the extremal RNBH.

On the other hand, we could inspect the violation of weak energy condition (WEC) for the hairy RNBHs since $V(\phi)$ is not entirely positive definite with $V(\phi) < 0$ in some regions of $\phi$,  thus the expression of WEC is given by
\begin{equation} \label{WEC}
 \rho = - T^t_{t} = \frac{N}{2} \phi'^2 + V  + \frac{Q^2}{2 r^4} \,.
\end{equation}
We observe that $\rho$ approaches zero at the infinity but at the horizon, given that $N(r_H)=0$, hence the WEC of neutral hairy RNBHs which is from our previous work \cite{Chew:2022enh} is violated at the horizon with $\rho=V(\phi_H)<0$. Therefore, it will be interested to study could the inclusion of $Q$ reduce the violation of WEC.

\section{Results and Discussions}\label{sec:res}

Recall that the input parameters in the calculation are given by $\phi_0$, $\phi_1$, $r_H$, $q$ and $\phi_H$. To generate the solutions of charged hairy black holes, we choose the values of global minimum $\phi_1$ as $\phi_1=0.5, 1.0$. For each $\phi_1$, we fix several values for the electrical charge $q$ in the range $[0,1]$ and then increase the scalar field at the horizon $\phi_H$ from $\phi_H=0$ until $\phi_H=\phi_1$, hence we find that a family of charged hairy black holes bifurcates from the RNBH when $\phi_H$ is non-trivial. Here we identify the charged hairy black holes as the hairy RNBHs where the non-extremal case corresponds to $0\leq q < 1$ and the extremal case corresponds to $q=1$. Subsequently, we present their properties based on our numerical results. 

\begin{figure}
\centering
\mbox{
(a)
 \includegraphics[angle =-90,scale=0.33]{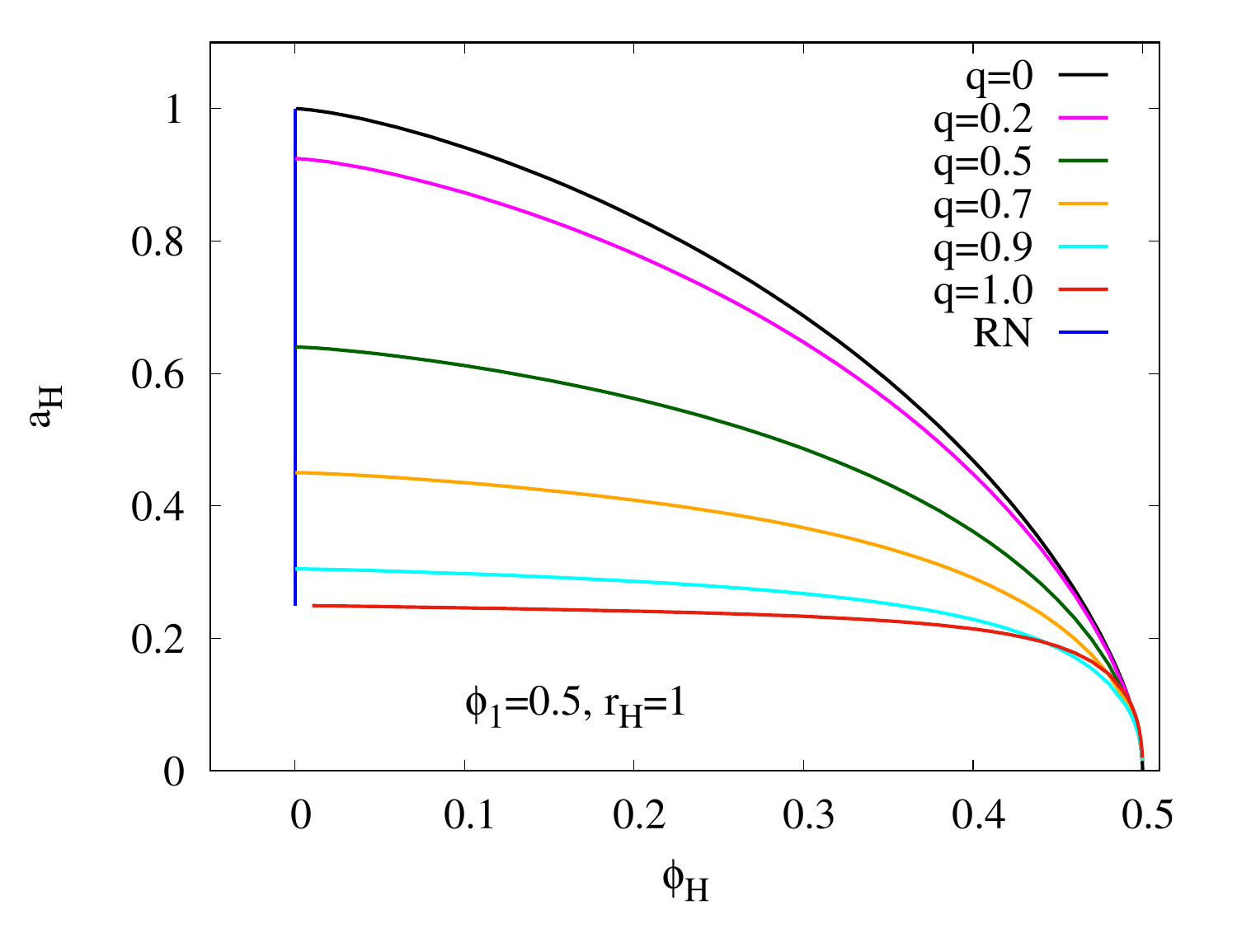}
(b)
 \includegraphics[angle =-90,scale=0.33]{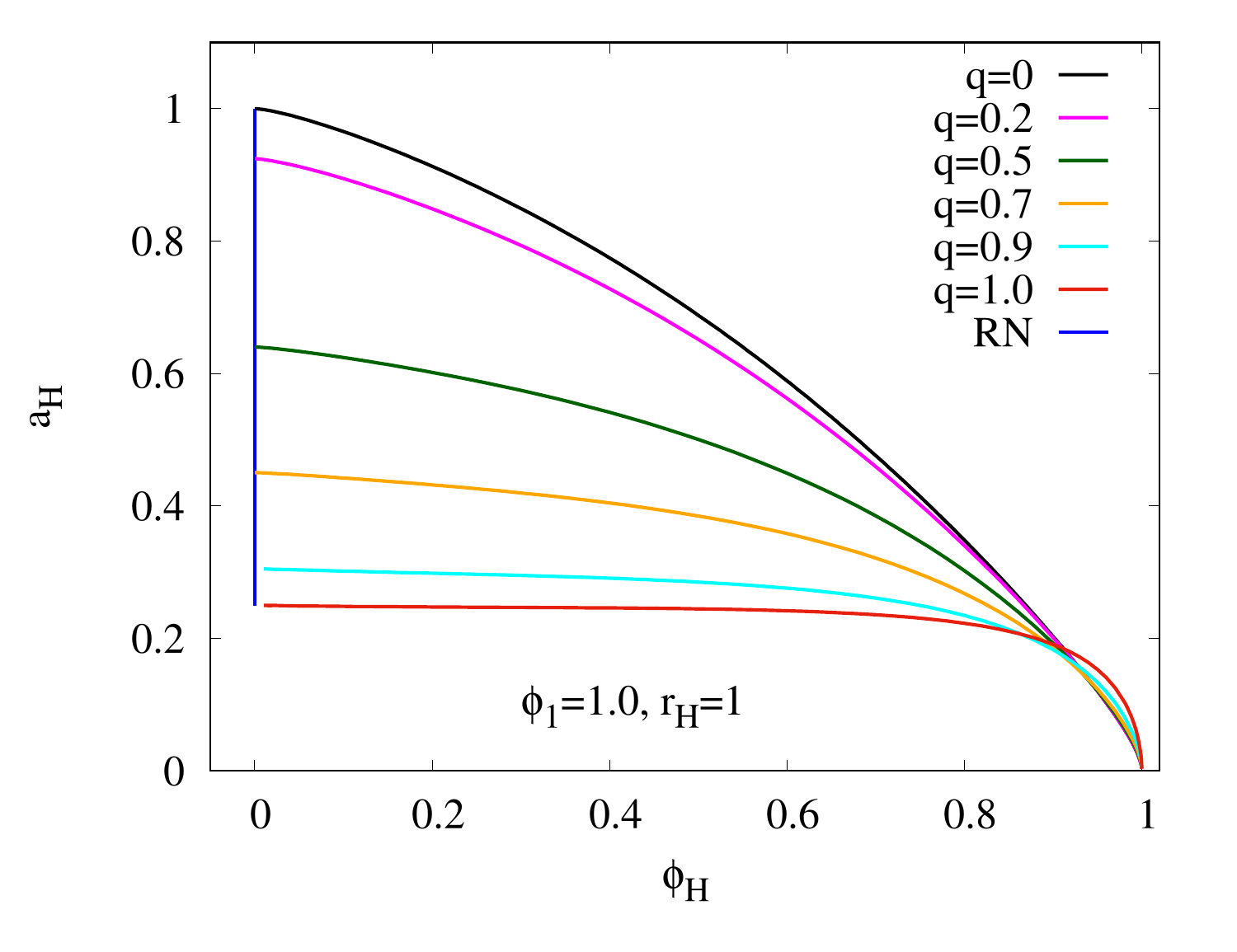}
 }
\caption{
The reduced area of horizon $a_H$ of the hairy RNBHs with $r_H=1$ and several $q$ for (a) $\phi_1=0.5$, (b) $\phi_1=1.0$. 
}
\label{plot_aH}
\end{figure}

First, we exhibit the reduced area of horizon $a_H$ for the hairy RNBHs in Fig.~\ref{plot_aH} with (a) $\phi_1=0.5$ and (b) $\phi_1=1.0$. When $\phi_H=0$, the charged black hole is merely the RNBH, and $a_H$ is bounded in between $[0.25,1]$ (blue curve) where $a_H=0.25$ corresponds to the extremal RNBH $(q=1)$ and $a_H=1$ corresponds to the Schwarzschild black hole $(q=0)$. When $q=0$, a branch of neutral hairy black holes (black curve) bifurcates from the Schwarzschild black hole when $\phi_H$ increases from $\phi_H=0$ to $\phi_H=\phi_1$ where their properties have been studied extensively by the authors in \cite{Chew:2022enh}. When $q>0$, we also increase $\phi_H$ from $\phi_H=0$ until $\phi_H=\phi_1$, a family of hairy RNBHs bifurcates from the RNBH where hairy RNBHs behave very differently than the RNBH, we find that $a_H$ decreases monotonically from unity to zero. In the limit $\phi_H=\phi_1$, $\phi_H$ sits exactly at the global minimum of $V(\phi)$, hence hairy RNBHs do not exist anymore. Overall, $a_H$ for both cases $\phi_1=0.5$ and $\phi_1=1.0$ behave qualitatively the same. Note that for a fixed value of $\phi_H$, $a_H$ decreases with the increase of $q$, which implies the ADM mass of hairy RNBHs increases when $q$ increases, thus the extremal hairy RNBHs (red curve) are the most massive hairy black holes while the neutral hairy black holes (black curve) are the lightest hairy black holes. Moreover, the value of $q$ in our theory is bounded for $0 \leq q \leq 1$, where this is in contrast to the charged hairy black holes in the Einstein-Maxwell-scalar case where they can possess $q$ greater than 1 \cite{Astefanesei:2019pfq,Kiorpelidi:2023jjw}, thus our hairy RNBHs are not overcharged.   

\begin{figure}
\centering
\mbox{
(a)
 \includegraphics[angle =-90,scale=0.33]{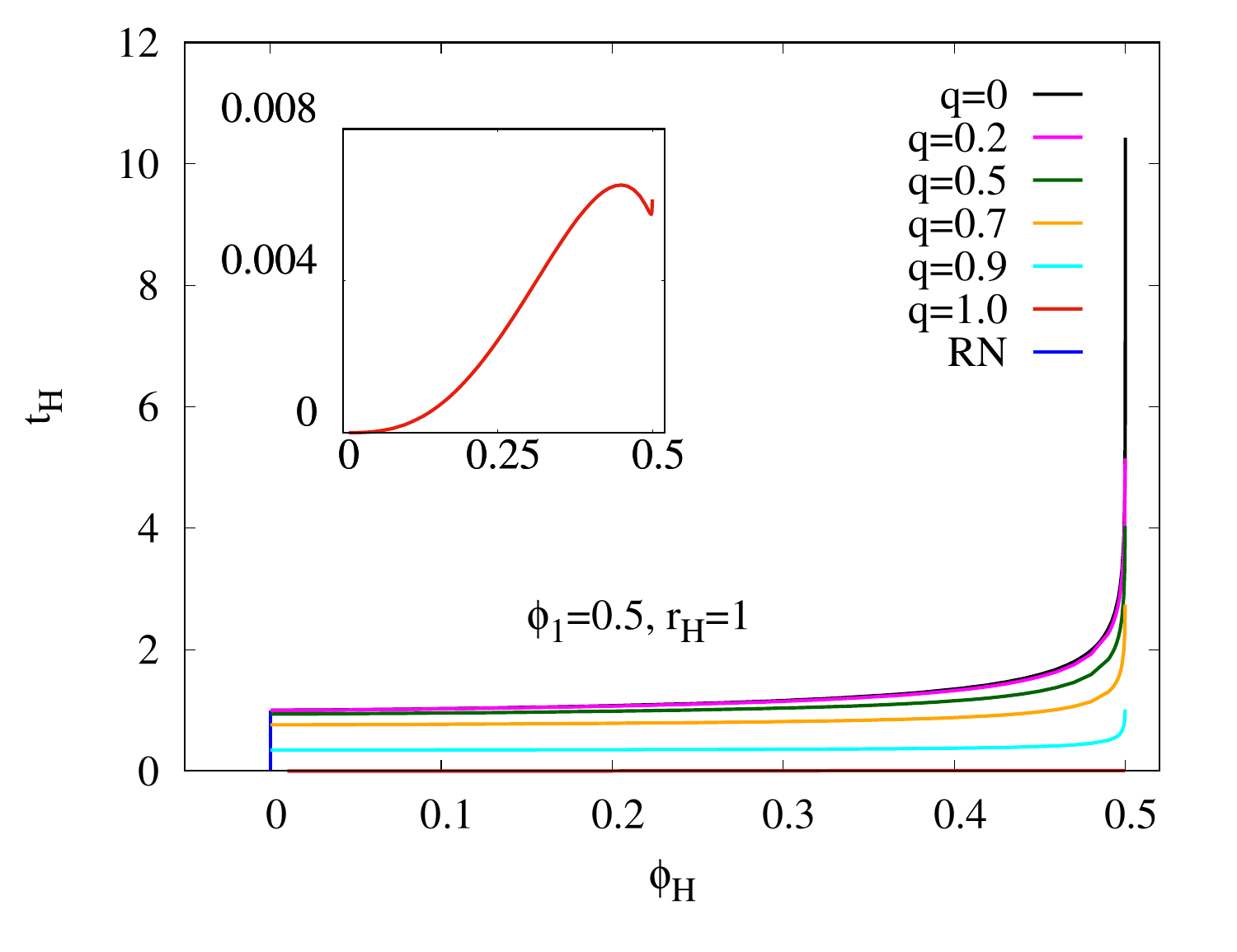}
(b)
 \includegraphics[angle =-90,scale=0.33]{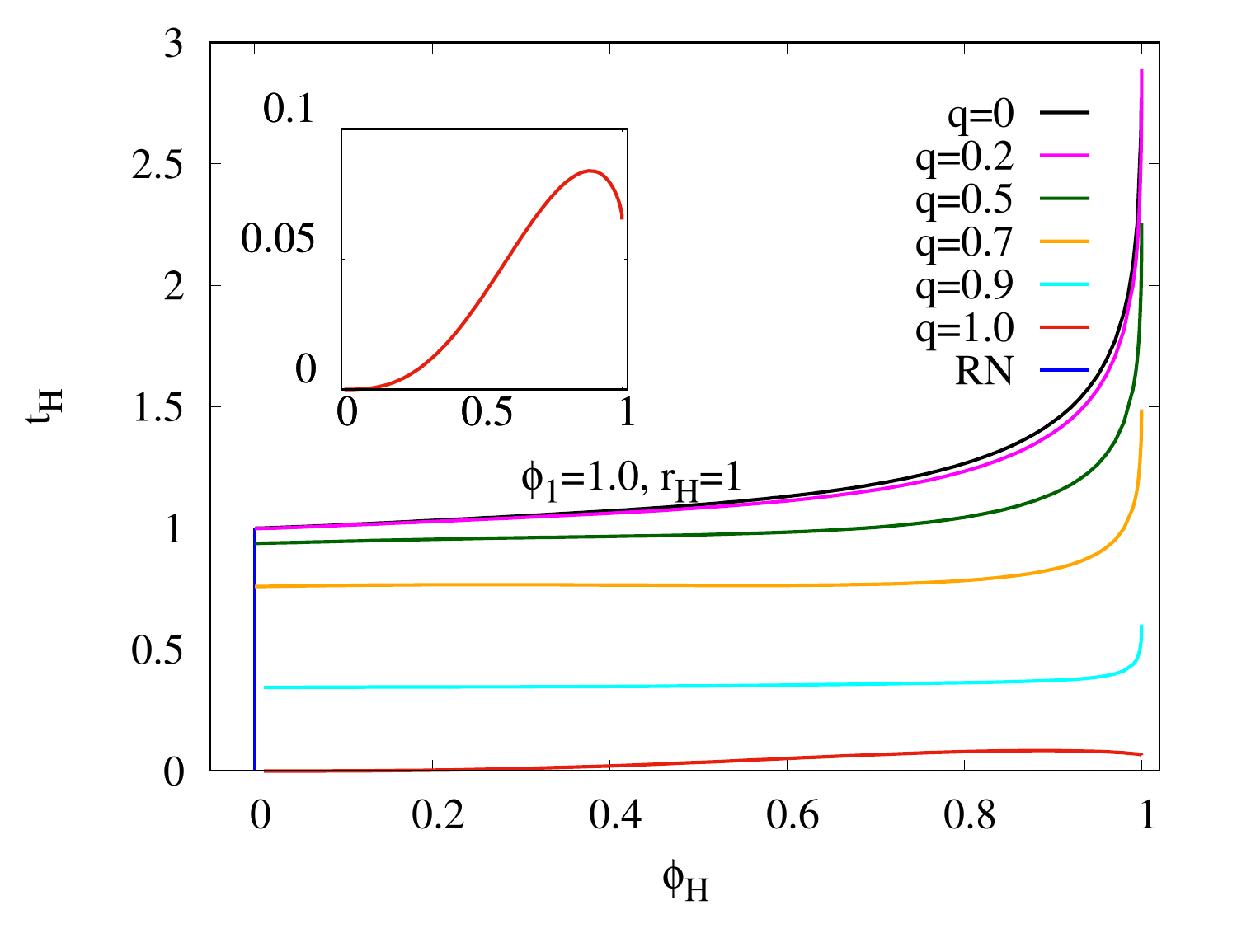}
 }
\caption{
The reduced Hawking temperature $t_H$ of the hairy RNBHs with $r_H=1$ and several $q$ for (a) $\phi_1=0.5$, (b) $\phi_1=1.0$. 
}
\label{plot_tH}
\end{figure}

Second, we show the reduced Hawking temperature $t_H$ for the hairy RNBHs in Fig.~\ref{plot_tH} with (a) $\phi_1=0.5$ and (b) $\phi_1=0.5$. Analogous to $a_H$, when $\phi_H=0$, $t_H$ (blue curve) is bounded in $0 \leq t_H \leq 1$ where $t_H=1$ corresponds to the Schwarzcshild black hole $(q=0)$ and $t_H=0$ corresponds to the extremal Reissner-Nordstrom black hole $(q=1)$. Similarly, when $\phi_H$ increases, a branch of hairy RNBHs emerges from the RNBH for a fixed value of $q$. When $q=0$, the corresponding hairy RNBHs are neutral where they have been considered by the authors in Ref.~\cite{Chew:2022enh}. When $0 < q <1$, $t_H$ increases very sharply when $\phi_H$ increases from $\phi_H=0$ to $\phi_H=\phi_1$, this also indicates that the hairy RNBHs don't exist when $\phi_H=\phi_1$. Interestingly, the inset of Figs.~\ref{plot_tH}(a) and (b) demonstrate that $t_H$ possesses a non-zero value when $q=1$ where the extremal hairy RNBH emerges from the extremal RNBH. Furthermore, $t_H$ behaves qualitatively the same for both cases $\phi_1=0.5$ and $\phi_1=1.0$.    

\begin{figure}
\centering
\mbox{
(a)
 \includegraphics[angle =-90,scale=0.33]{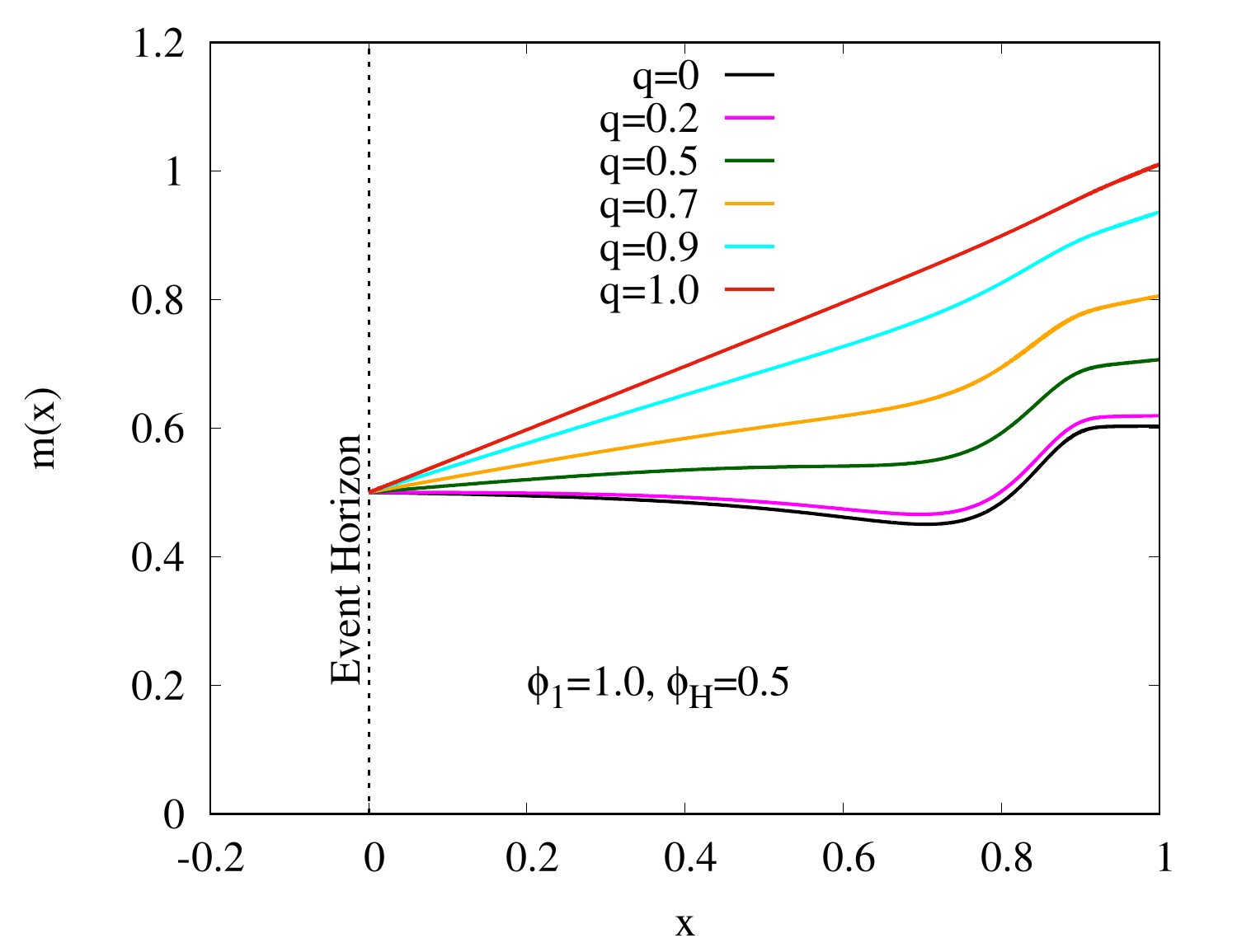}
(b)
 \includegraphics[angle =-90,scale=0.33]{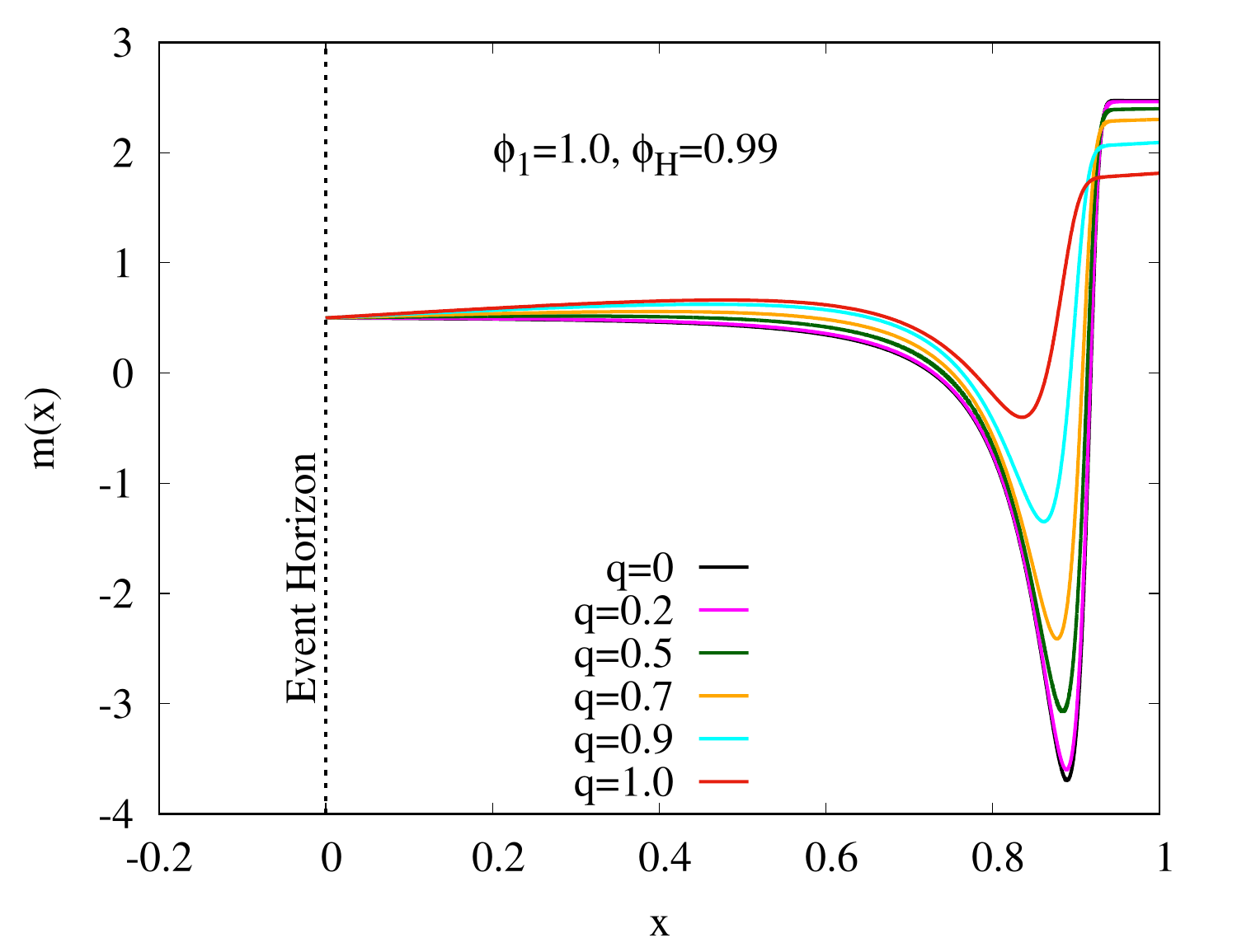}
}
\caption{
The profiles of mass function $m(x)$ in the compactified coordinate $x$ for the hairy RNBHs with $r_H=1$, $\phi_1=1.0$ and several $q$ for (a) $\phi_H=0.5$, (b) $\phi_H=0.99$.
}
\label{plot_m}
\end{figure}

Fig.~\ref{plot_m} exhibits the profiles of mass function $m(x)$ of the hairy RNBHs with $r_H=1$ and $\phi_1=1.0$ in the compactified coordinate $x$ for (a) $\phi_H=0.5$ and (b) $\phi_H=0.99$ $(\text{in the limit} \, \phi_H \rightarrow \phi_1)$. We observe that  $m(x)$ (black curve) with $q=0$ as depicted in Figs.~\ref{plot_m}(a) and (b) possess almost a constant function inside the bulk, corresponding to the global minimum of the potential $V(\phi)$ which is the true vacuum $\phi_1$. Moving away from the horizon, they develop a sharp boundary which looks like a global minimum at some intermediate region of the spacetime, where the functions rapidly change to another set of almost constant function which corresponds to the imposed false vacuum $(a=0)$ at infinity, where the scalar field sits in the local minimum. However, when $q$ increases as shown in Fig.~\ref{plot_m}(a) for $\phi_H=0.5$, the gradient of $m(x)$ at the horizon and the infinity also increase, thus $m(x)$ no longer possess almost constant functions inside the bulk and at the infinity, this has reduced the sharp boundary in some intermediate region of the spacetime, as the consequence the global minimum is being lifted and then finally disappears. Thus, $m(x)$ (red curve) becomes almost a linear function when $q=1$ for extremal hairy RNBHs. Nevertheless, Fig.~\ref{plot_m}(b) demonstrates that when $\phi_H=0.99$, $m(x)$ still possess a sharp boundary that connects two different sets of almost constant functions at the horizon and the infinity, although that sharp boundary has been reduced with the increasing of $q$, therefore the global minimum of $m(x)$ still can exist even for $q=1$ (red curve). Meanwhile, we find a very interesting phenomena that the profiles of solutions are heavily dominated by either the strength of $q$ or $\phi_H$.

\begin{figure}
\centering
\mbox{
(a)
 \includegraphics[angle =-90,scale=0.33]{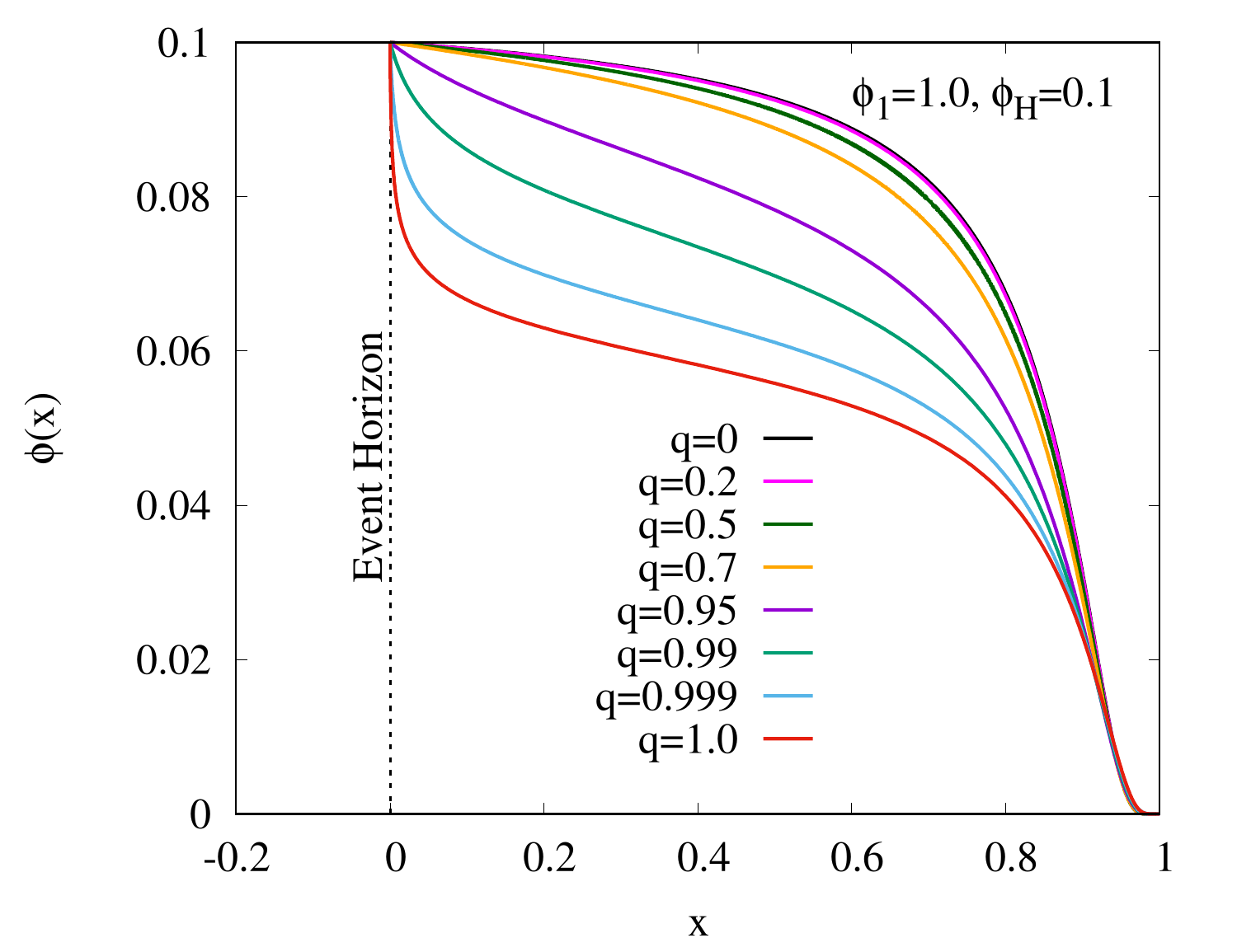}
(b)
 \includegraphics[angle =-90,scale=0.33]{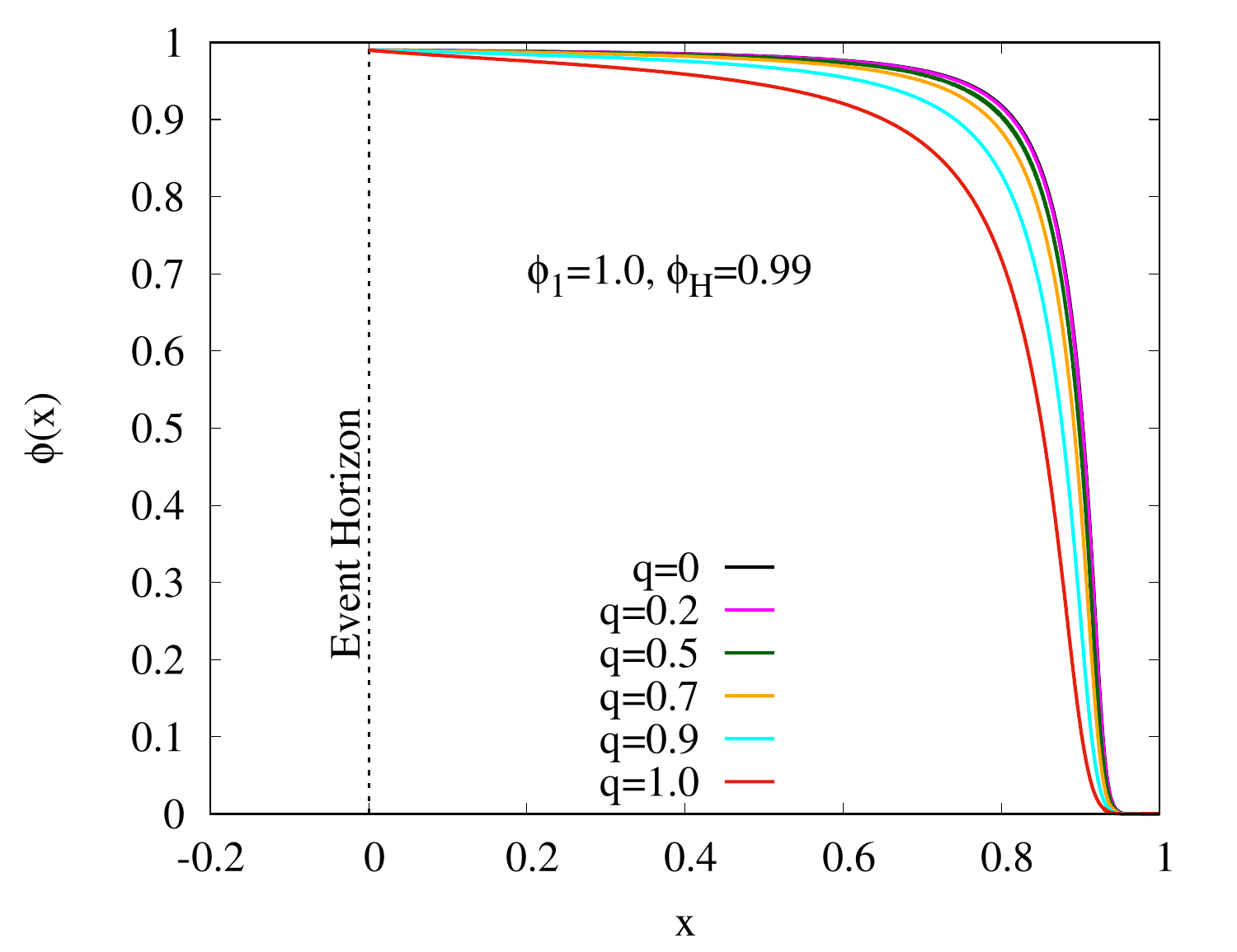}
}
\caption{
The profiles of scalar field $\phi(x)$ in the compactified coordinate $x$ for the hairy RNBHs with $r_H=1$, $\phi_1=1.0$ and several $q$ for (a) $\phi_H=0.1$, (b) $\phi_H=0.99$.
}
\label{plot_phi}
\end{figure}

Fig.~\ref{plot_phi} exhibits the profiles of scalar field $\phi(x)$ of the hairy RNBHs with $r_H=1$ and $\phi_1=1.0$ in the compactified coordinate $x$ for (a) $\phi_H=0.1$ and (b) $\phi_H=0.99$ $(\text{in the limit} \, \phi_H \rightarrow \phi_1)$. Overall $\phi(x)$ decreases monotonically to zero from the horizon to the infinity. As shown in Fig.~\ref{plot_phi}(a), when $\phi_H=0.1$, the behavior of $\phi(x)$ is qualitatively similar to $m(x)$ where initially the gradient of $\phi(x)$ is very small for small $q$ but it becomes the steepest at the horizon when $q=1$. This phenomenon can be described by the denominator of $\phi_{H,1}$ (Eq.~\eqref{hor}) where $V(\phi_H)<0$ and $\frac{dV(\phi_H)}{d\phi}<0$, the increasing of $q$ decreases the denominator of $\phi_{H,1}$, hence $\phi_{H,1}$ increases and then becomes largest but still remains finite when $q=1$. Nevertheless, when we increase $\phi_H$, for instance, $\phi_H=0.99$ as shown in Fig.~\ref{plot_phi}(b), the increasing of $\phi_H$ can flatten the profile of scalar field in the bulk because the increasing of $V(\phi_H)$ increases the denominator of $\phi_{H,1}$, hence $\phi(x)$ looks like an almost constant function and unaffected by $q$ in the bulk.

\begin{figure}
\centering
\mbox{
(a)
 \includegraphics[angle =-90,scale=0.33]{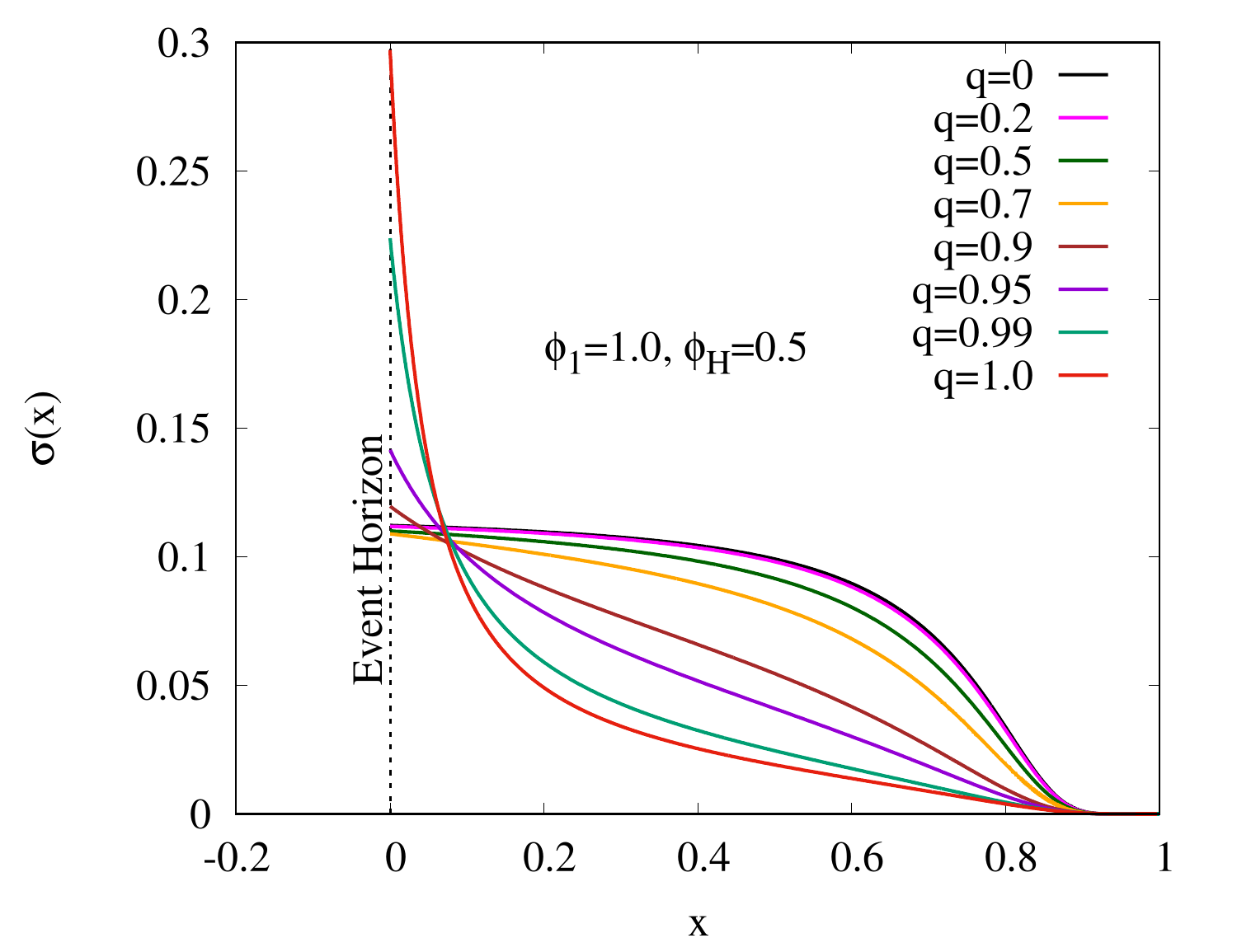}
(b)
 \includegraphics[angle =-90,scale=0.33]{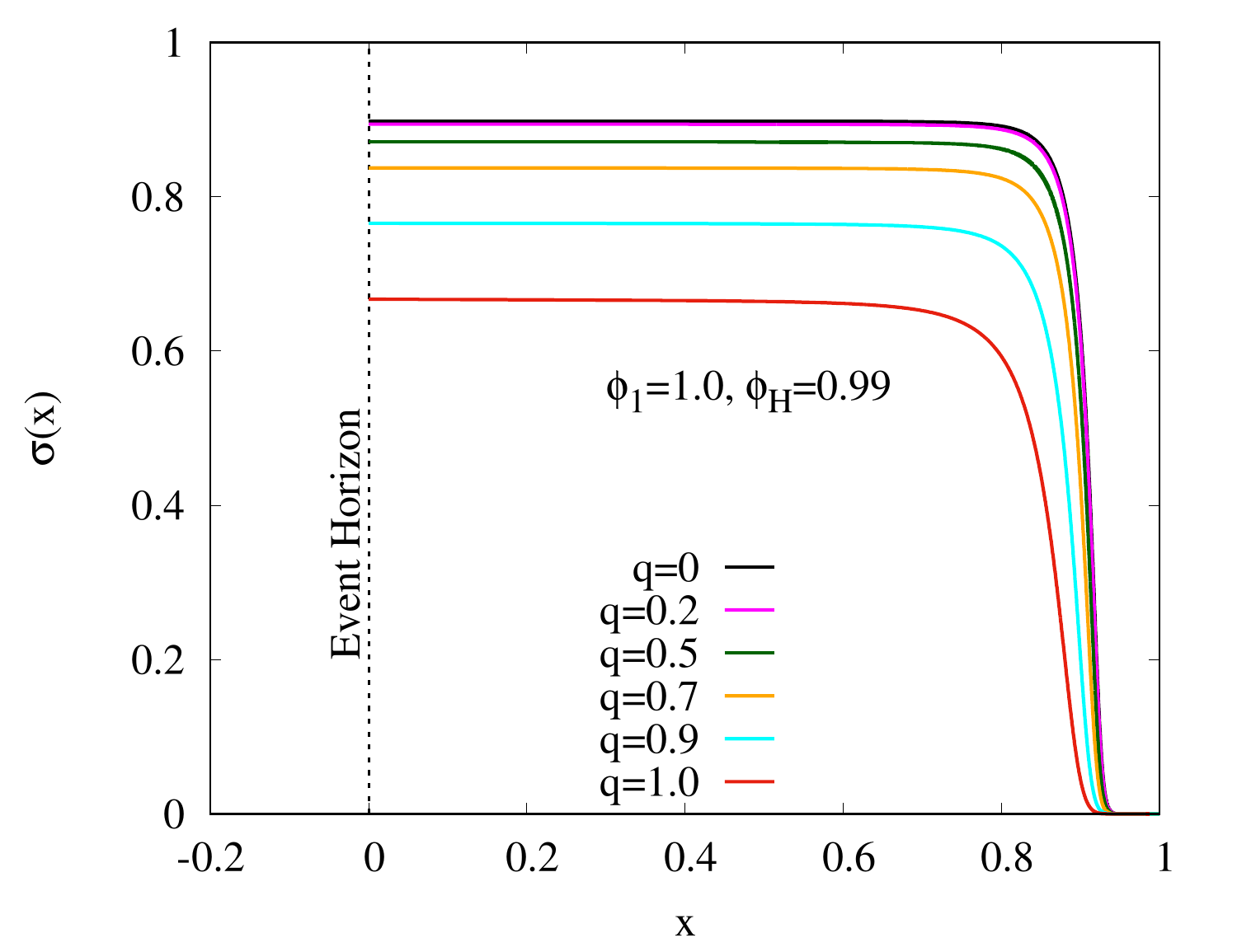}
}
\caption{
The profiles of metric function $\sigma(x)$ in the compactified coordinate $x$ for the hairy RNBHs with $r_H=1$, $\phi_1=1.0$ and several $q$ for (a) $\phi_H=0.5$, (b) $\phi_H=0.99$.
} 
\label{plot_sigma}
\end{figure}

Fig.~\ref{plot_sigma} exhibits the profiles of metric function $\sigma(x)$ of the hairy RNBHs with $r_H=1$ and $\phi_1=1.0$ in the compactified coordinate $x$ for (a) $\phi_H=0.5$ and (b) $\phi_H=0.99$ $(\text{in the limit} \, \phi_H \rightarrow \phi_1)$. In general $\sigma(x)$ behaves quite similar with the function $\phi(x)$ where it also decreases monotonically to zero from the horizon to the infinity. As shown in Fig.~\ref{plot_sigma}(a) for $\phi_H=0.1$, the profile of $\sigma(x)$ is analogous to $\phi(x)$ for small $q$ where the gradient of $\sigma(x)$ is very small which also looks like almost a constant function at the horizon that corresponds to the true vacuum $\phi_1$. However, the value of $\sigma(x)$ and its gradient at the horizon increase very sharply when $q$ increases where it follows the similar changes of $\phi(x)$ since $\sigma_1 \propto \phi^2_{H,1}$ (see Eq.~\eqref{hor}). Hence, $\sigma(x)$ (red curve) is the steepest at the horizon when $q=1$. However, when $\phi_H$ is large, for instance as demonstrated in Fig.~\ref{plot_sigma}(b), the gradient of $\sigma(x)$ at the horizon is unaffected by any values of $q$ since $\phi_{H,1}$ is very small, thus we observe that $\sigma(x)$ still can behave like an almost constant function in the bulk. 

\begin{figure}
\centering
\mbox{
(a)
 \includegraphics[angle =-90,scale=0.33]{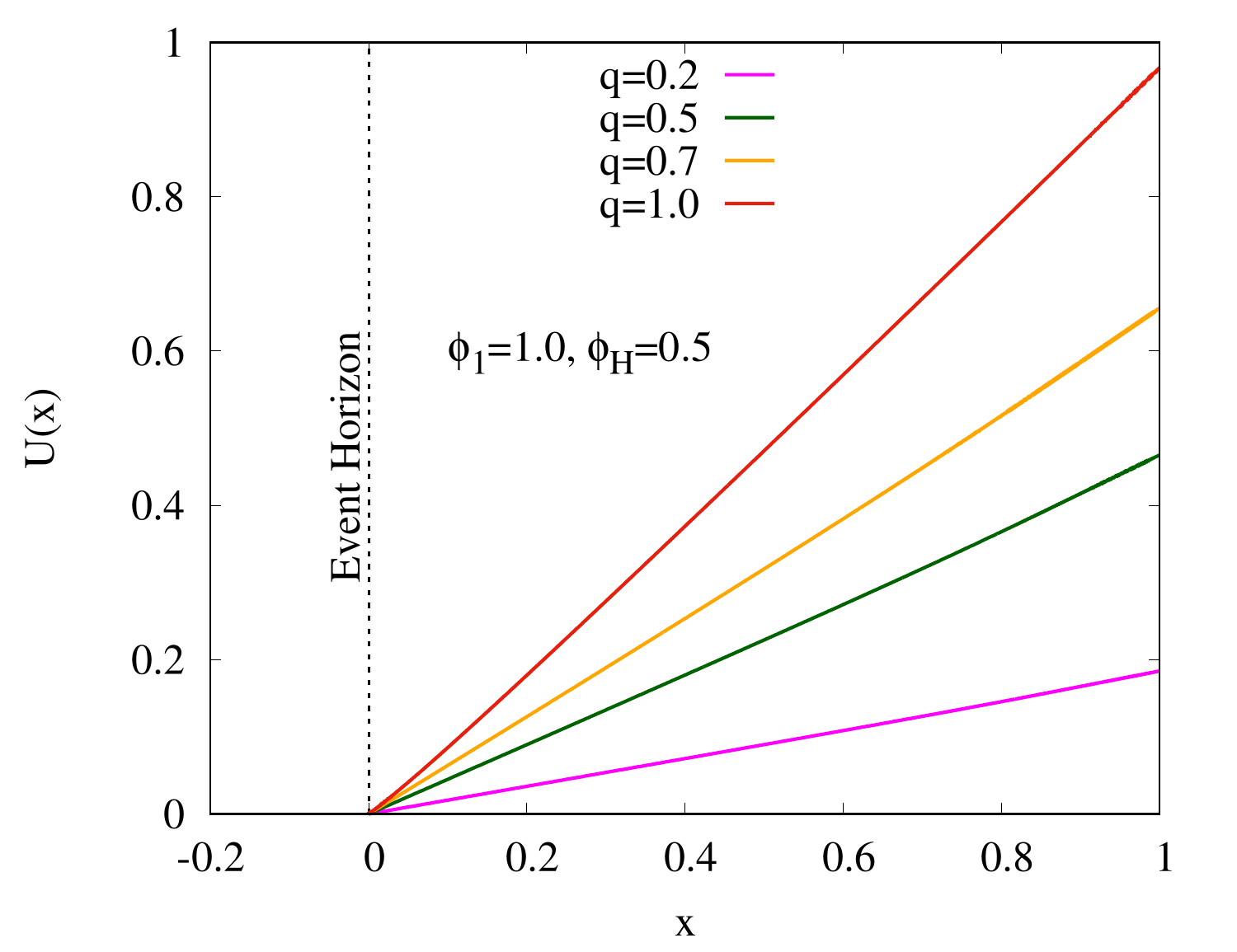}
(b)
 \includegraphics[angle =-90,scale=0.33]{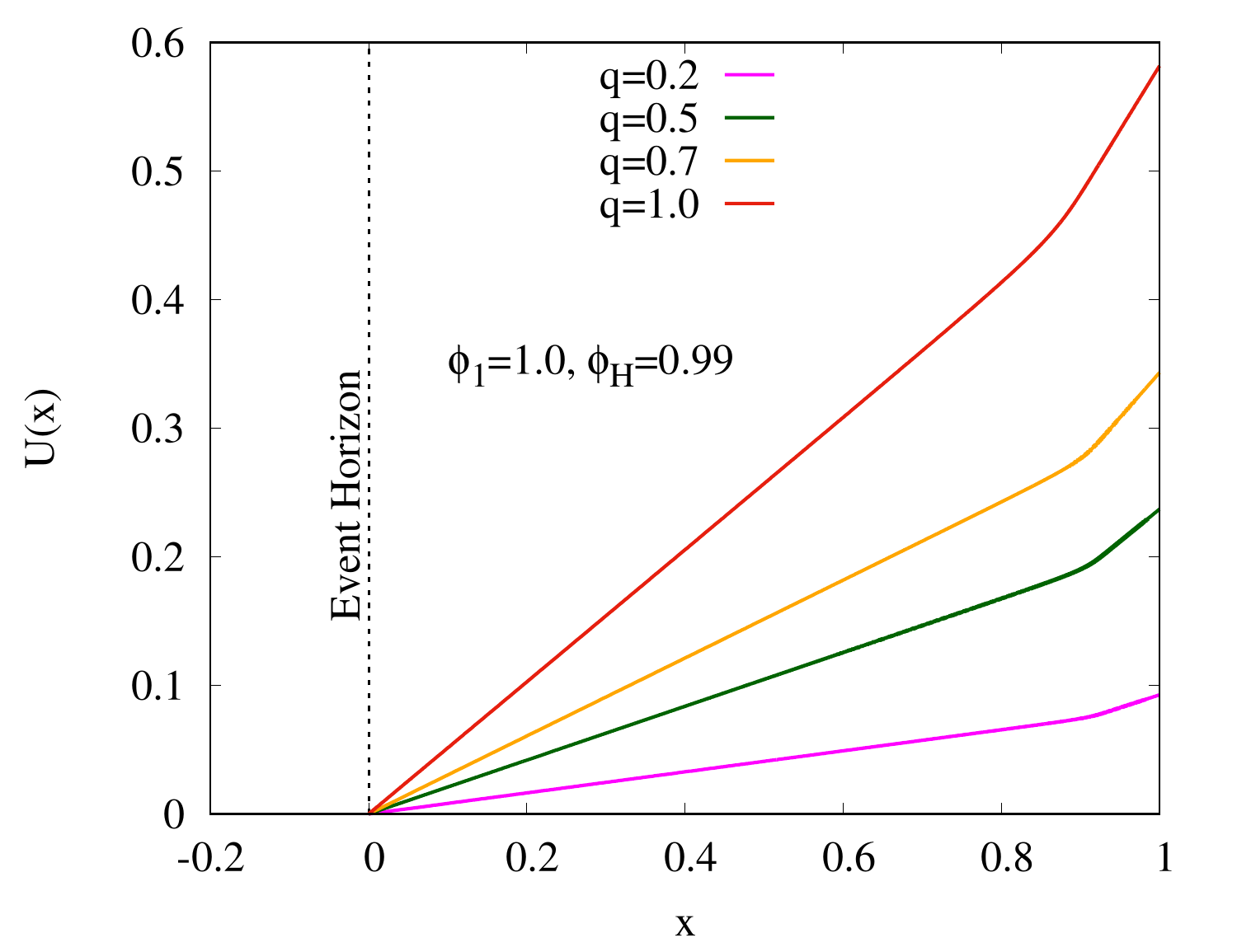}
}
\caption{
The profiles of gauge field $U(x)$ in the compactified coordinate $x$ for the hairy RNBHs with $r_H=1$, $\phi_1=1.0$ and several $q$ for (a) $\phi_H=0.5$, (b) $\phi_H=0.99$.
}
\label{plot_U}
\end{figure}

Fig.~\ref{plot_U} shows the profiles of gauge field $U(x)$ of the hairy RNBHs with $r_H=1$ and $\phi_1=1.0$ in the compactified coordinate $x$ for (a) $\phi_H=0.5$ and (b) $\phi_H=0.99$. In Fig.~\ref{plot_U}(a), the profile of $U(x)$ is a linear function, and its gradient increases with the increase of $q$. In Fig.~\ref{plot_U}(b), $U(x)$ is still a linear function that increases linearly with the increase of $x$ but increases a little sharply with a larger gradient near $x=1$.  

\begin{figure}
\centering
\mbox{
(a)
 \includegraphics[angle =-90,scale=0.33]{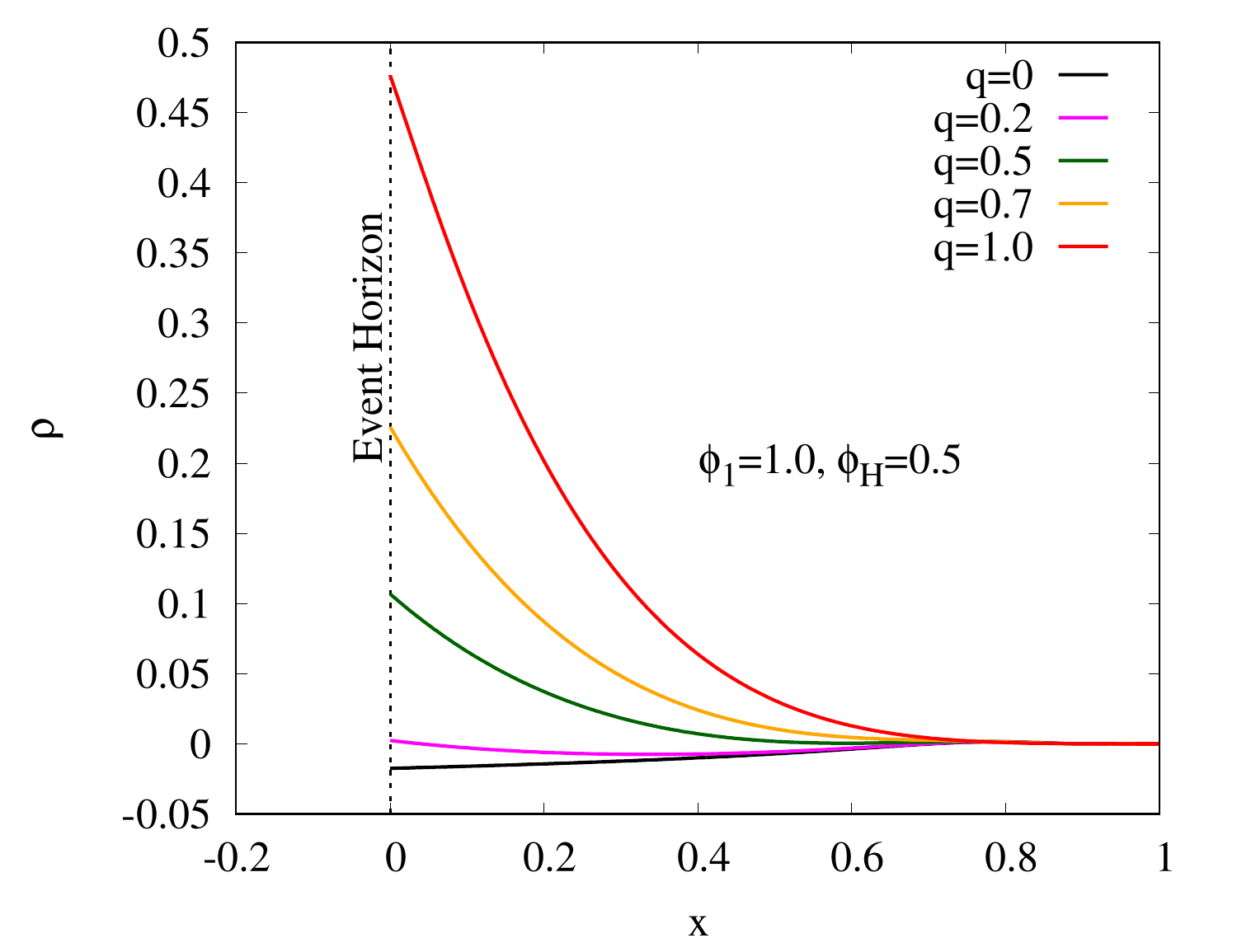}
(b)
 \includegraphics[angle =-90,scale=0.33]{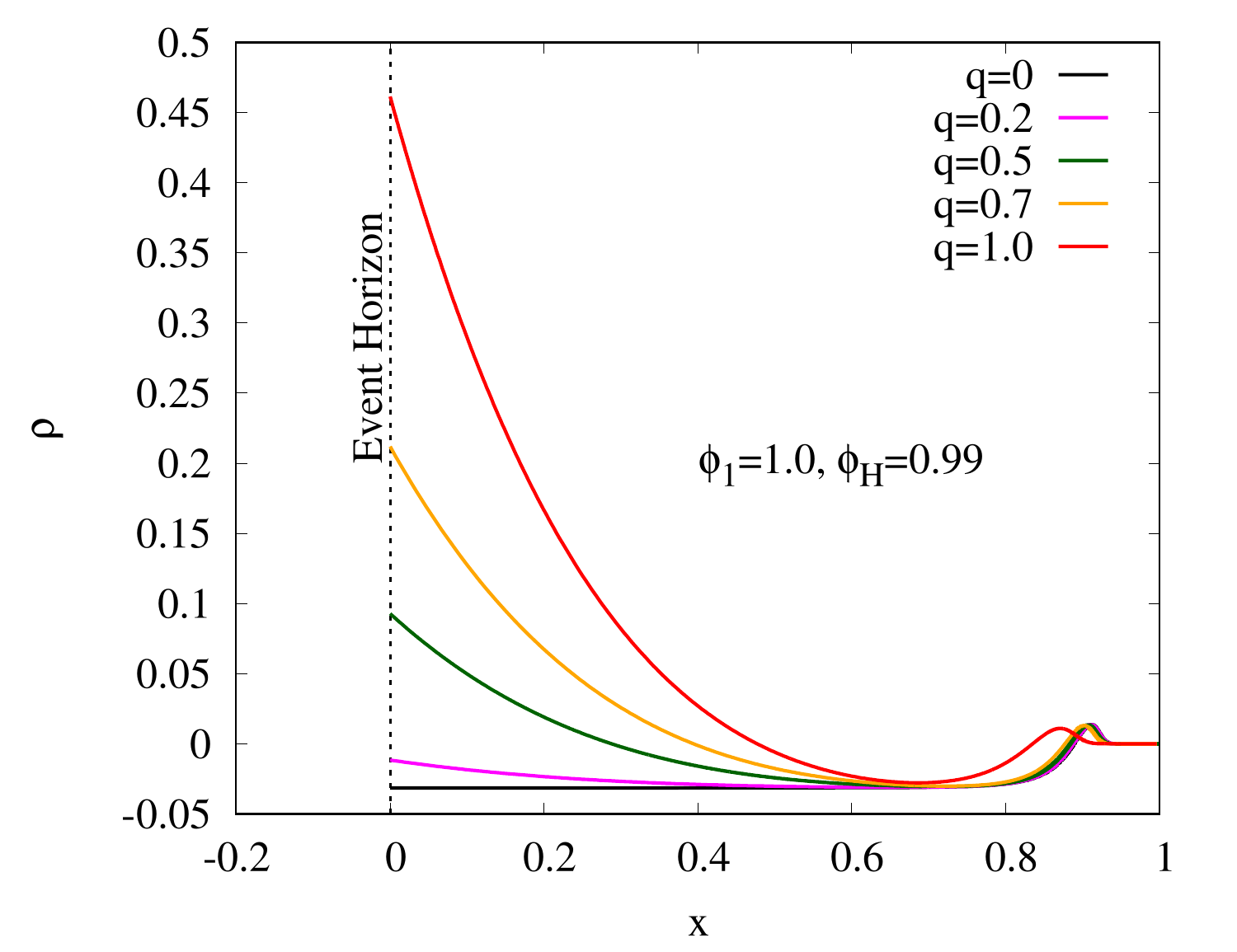}
 }
\caption{
The weak energy condition $\rho$ in the compactified coordinate $x$ for the hairy RNBHs with $r_H=1$, $\phi_1=1.0$ and several $q$ for (a) $\phi_H=0.5$ and (b) $\phi_H=0.99$.
}
\label{plot_Ttt}
\end{figure}

Fig.~\ref{plot_Ttt} shows the weak energy condition (WEC) of Eq.~\eqref{WEC} with several values of $q$ in the compactified coordinate $x$ for (a) $\phi_H=0.5$ and $\phi_H=0.99$. When $\phi_H=0.5$ and $q=0$, we observe that the WEC is violated since the local energy density is negative $(\rho=V(\phi_H)<0)$, particularly at the horizon. When we increase $q$, $\rho$ can become strictly positive since the inclusion of $q$ can reduce the violation of WEC. Hence, it gives us an important hint that the WEC of neutral hairy black holes in Ref.~\cite{Chew:2023olq} can be possibly satisfied if they become charged black holes. When $\phi_H=1.0$ and $q=0$, WEC is violated at the horizon since $\rho<0$. Similarly, WEC is also being satisfied at the horizon when we increase $q$. However, WEC is slightly violated in some regimes of $x$ since $\rho<0$. Note that there are some small peaks that form near $x=1$ due to the global minimum of $m(x)$.

\begin{figure}
\centering
\mbox{
(a)
 \includegraphics[angle =-90,scale=0.33]{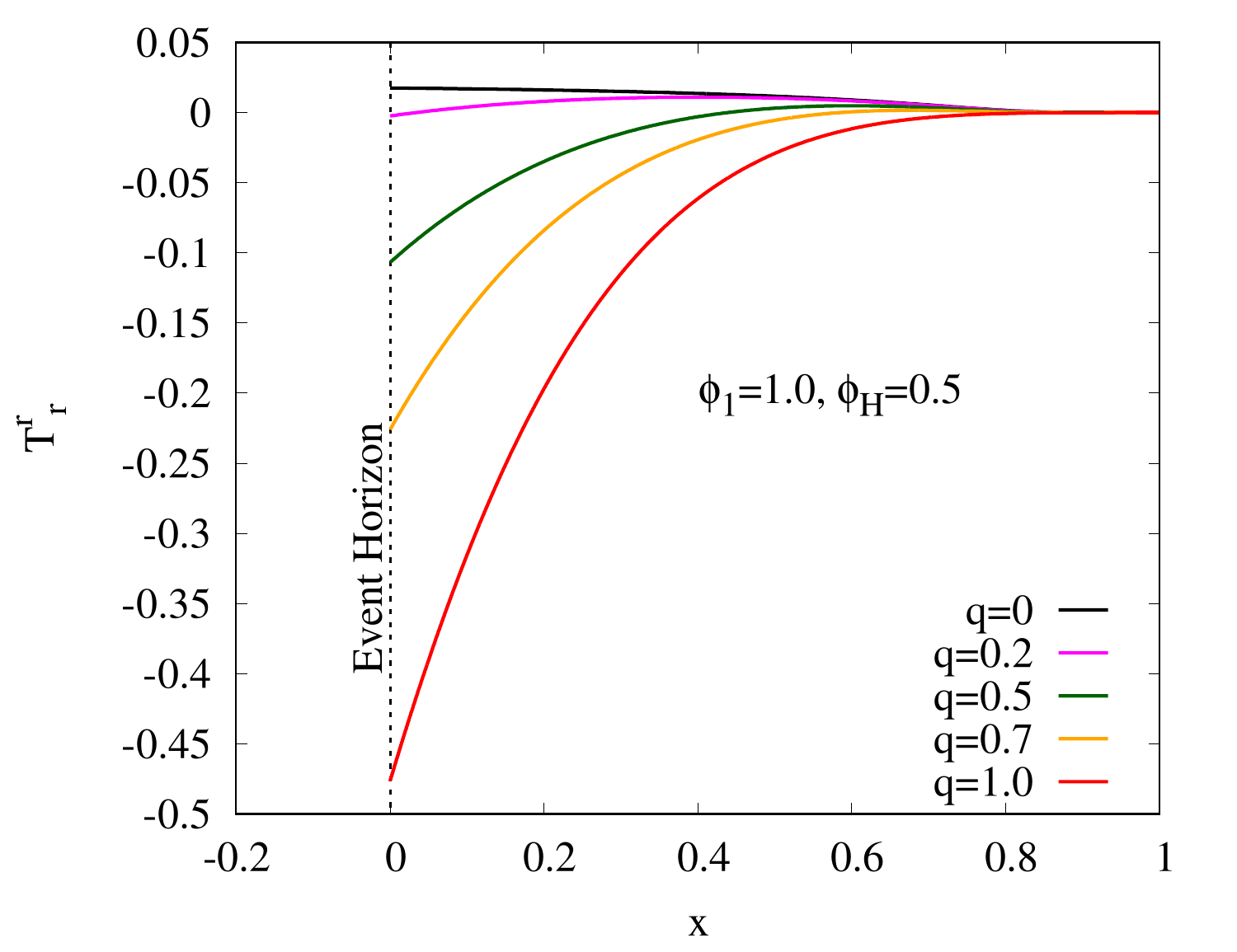}
(b)
 \includegraphics[angle =-90,scale=0.33]{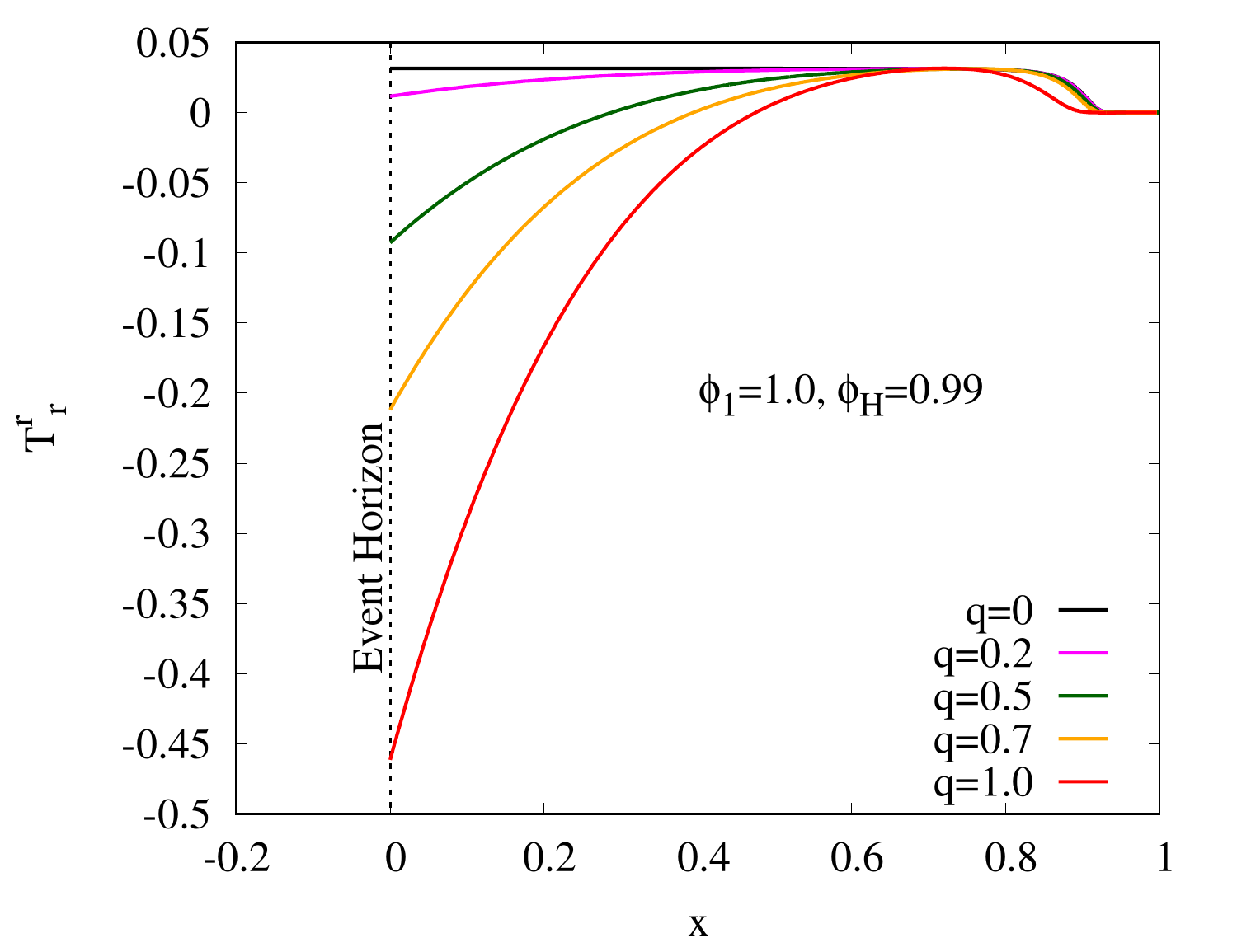}
 }
\caption{
The energy condition $T^r\,_r$ in the compactified coordinate $x$ for the hairy RNBHs with $r_H=1$, $\phi_1=1.0$ and several $q$ for (a) $\phi_H=0.5$ and (b) $\phi_1=0.99$.
}
\label{plot_Trr}
\end{figure}

Meanwhile, the energy condition of $T^r_r$ is depicted in Fig.~\ref{plot_Trr} for a) $\phi_H=0.5$ and b) $\phi_H=0.99$. When $\phi_H=0.5$ and $q=0$, the energy condition of $T^r_r$ is satisfied since $T^r_r>0$. However, when we increase $q$, $T^r_r<0$, particularly at the horizon, hence its energy condition is being violated. Similarly, when $\phi_H=0.99$ and $q=0$, the energy condition of $T^r_r$ is being satisfied since it is strictly positive. Nevertheless, when we increase $q$, $T^r_r<0$, particularly at the horizon, hence its energy condition is being violated. Analogous to $\rho$, $T^r_r$ possesses a local maximum exactly at the location of the global minimum of $m(x)$.

\section{Conclusion}\label{sec:con}

In this paper, we have broadened the class of hairy Reissner-Nordstrom black holes (RNBH) by minimally coupling a scalar potential $V(\phi)$ in the Einstein-Maxwell-Klein-Gordon (EMKG) theory to construct charged hairy black holes. Here we introduce $V(\phi)$ which possesses a local minimum, a global minimum $\phi_1$, and a local maximum $\phi_0$ where it has been applied in the cosmology to describe the phase transition of bubbles from the false vacuum (local maximum) to the true vacuum (global minimum). Previously the corresponding $V(\phi)$ has been applied by the authors to construct the asymptotically flat neutral hairy black holes \cite{Chew:2022enh}, and hence, we generalize them to the charged hairy black holes which possess the electric charge $Q$. 

In the EMKG theory, the trivial solution is the RNBH with the scalar field diverges at the horizon when $V(\phi)$ does not exist. When we include $V(\phi)$, we demonstrate that it is possible to circumvent the no-hair theorem where we can obtain the charged hairy black holes that are globally regular outside the horizon. Thus, the family of hairy charged black holes emerges from the RNBH with a fixed value of charge per unit mass, $q=Q/M$ when the scalar field at the horizon $\phi_H$ is non-trivial and we can identify the charged hairy black holes as the hairy RNBHs. The value of $q$ is bounded in $[0,1]$ where $q=0$ corresponds to the neutral hairy black holes \cite{Chew:2022enh}, $0\leq q<1$ corresponds to non-extremal hairy RNBHs, while $q=1$ corresponds to the extremal hairy RNBHs. This is also in contrast to the hairy black holes in Einstein-Maxwell-scalar theory where they can possess $q>1$ \cite{Astefanesei:2019pfq,Kiorpelidi:2023jjw}.

The properties of neutral hairy black holes $(q=0)$ have been studied previously by the authors \cite{Chew:2022enh}. Here we study the properties of hairy RNBHs by choosing $\phi_1=0.5, 1.0$, then we fix a value of $q$, when we increase $\phi_H$ from zero, we find that the reduced area of horizon $a_H$ decreases from unity to zero while the reduced Hawking temperature $t_H$ increases very sharply when $\phi_H=\phi_1$. This might imply that the hairy RNBHs don't exist in that limit. Note that the extremal hairy RNBHs possess the non-trivial $t_H$ while RNBH possesses the vanishing $t_H$. 

Then we briefly summarize the profiles of solutions. When $q=0$, the mass function possesses almost a constant function inside the bulk, corresponding to the global minimum of the potential $V(\phi)$ which is the true vacuum $\phi_1$. Moving away from the horizon, it develops a sharp boundary which looks like a global minimum at some intermediate region of the spacetime, where the function rapidly change to another set of almost constant function which corresponds to the imposed false vacuum $(a=0)$ at infinity, where the scalar field sits in the local minimum. When $\phi_H=0.5$ with $q \neq 0$, the gradient of mass function at the horizon and the infinity increases, hence the sharp boundary is reduced and then the global minimum is being lifted when we increase $q$, thus the global minimum disappears for the extremal case $(q=1)$, the mass function looks like a linear function as a consequence. When $\phi_H=0.99$, the two different sets of almost constant functions are still connected by a sharp boundary although the global minimum is also being lifted slightly but eventually still can be preserved when $q=1$. The scalar field decreases monotonically from its maximum value at the horizon to zero at the infinity. When $\phi_H=0.1$, the scalar field also possesses an almost constant function at the horizon but its gradient at the horizon increases very sharply when $q$ increases, thus it is the steepest at the horizon when $q=1$. However, the gradient of the scalar field at the horizon can still become very small when $\phi_H=0.99$ even $q=1$. Furthermore, the gauge field increases linearly from the horizon to the infinity. When $\phi_H=0.99$, the gradient of the gauge field slightly increases near the infinity for $\phi_1=1.0$. Overall, we find a very interesting phenomenon that the profiles of solutions are heavily dominated by either the electric charge $q$ or the scalar field.  

The weak energy condition (WEC) which can be described by the local energy density $\rho=-T^t\,_t$ is violated, particularly at the horizon for the neutral hairy RNBHs $(q=0)$. However, WEC can be satisfied with $\rho>0$ at the horizon when $q \neq 0$ for $\phi_1=0.5$. Thus, this could imply that the WEC of neutral hairy black holes \cite{Chew:2023olq} can be possibly satisfied if they become charged black holes. Nevertheless, WEC is slightly violated in the region of compactified coordinate $0.5 < x < 1$ for $\phi_1=1.0$ although $q$ increases. Meanwhile, the component of stress-energy tensor $T^r\,_r$ is strictly positive for neutral hairy black holes $(q=0)$ but becomes negative, particularly at the horizon when $q \neq 0$ for $\phi_1=0.5, 1.0$.  

Here we briefly comment on the linear stability of the hairy RNBHs. The neutral hairy black holes from our previous work \cite{Chew:2022enh} are found to be unstable against the linear perturbation. Hence, the solutions of the hairy RNBHs are also likely to be unstable against the linear perturbation but it would be interesting to perform a radial perturbation on the background solutions in the mode analysis to investigate could the presence of $q$ stabilize the configuration of hairy RNBHs.  

There also could be several extensions from this work. First, we can consider the dyon case where the hairy RNBHs can possess the magnetic charge and study their properties systematically. Then, it will be worthwhile to stress that the hairy RNBHs recently have gained some interest in the context of the Cauchy horizon theorems where there are some research works show that the inner Cauchy horizon may not exist if there is a scalar hair at the outer horizon \cite{Hansen:2014rua,Hansen:2015dxa,Nakonieczna:2015umf,Nakonieczna:2016iof,Brihaye:2016vkv}. Hence, the investigation of the dynamical evolution of charged hairy black holes will be useful for us to understand the process of formation of Cauchy horizon, since we lack the analytical proof on this \cite{Chew:2023upu}. 

\section*{Acknowledgement}
XYC is supported by the starting grant of Jiangsu University of Science and Technology (JUST). DY is supported by the National Research Foundation of Korea (Grant no.: 2021R1C1C1008622, 2021R1A4A5031460). XYC is grateful with the hospitality provided by the organizer of the conference ``String, Gravity and Cosmology 2023" which was held in Busan, Korea.

\end{document}